\documentclass[twocolumn,tighten]{aastex63}

\usepackage{times,natbib,graphicx,amsmath,multirow,xspace}

\newcommand{\nustar}{\textit{NuSTAR}\xspace}

\newcommand{\chandra}{\textit{Chandra}\xspace}
\newcommand{\nicer}{\textit{NICER}\xspace}

\newcommand{\rxte}{\textit{RXTE}\xspace}
\newcommand{\bep}{\textit{BeppoSAX}\xspace}
\newcommand{\xmm}{{\it XMM-Newton}\xspace}
\newcommand{\lc}{light-curve}
\newcommand{\ms}{$M_{\odot}$}

\newcommand{\lumcgs}{ergs~s$^{-1}$\xspace}
\newcommand{\rin}{$R_{\rm in}$}
\newcommand{\rg}{$R_{g}$}
\newcommand{\risco}{$R_{\mathrm{ISCO}}$}

\newcommand{\relxill}{{\sc relxill}}
\newcommand{\relxillns}{{\sc relxillNS}}
\newcommand{\source}{\mbox{4U~1735$-$44}\xspace}

\shorttitle{NICER-NuSTAR View of 4U 1735$-$44}
\shortauthors{Ludlam et al.}

\begin{document}

\title{\nicer--\nustar Observations of the Neutron Star Low-Mass X-ray Binary 4U 1735$-$44}

\correspondingauthor{R. M. Ludlam}
\email{rmludlam@caltech.edu}

\renewcommand{\thefootnote}{\fnsymbol{footnote}}

\author[0000-0002-8961-939X]{R.~M.~Ludlam}\thanks{NASA Einstein Fellow}
\affiliation{Cahill Center for Astronomy and Astrophysics, California Institute of Technology, Pasadena, CA 91125, USA}

\author[0000-0002-8294-9281]{E.~M.~Cackett}
\affiliation{Department of Physics \& Astronomy, Wayne State University, 666 West Hancock Street, Detroit, MI 48201, USA}

\author{J.~A.~Garc\'{i}a}
\affiliation{Cahill Center for Astronomy and Astrophysics, California Institute of Technology, Pasadena, CA 91125, USA}
\affiliation{Remeis Observatory \& ECAP, Universit\"{a}t Erlangen-N\"{u}rnberg, Sternwartstr. 7, D-96049, Bamberg, Germany}

\author{J.~M.~Miller}
\affiliation{Department of Astronomy, University of Michigan, 1085 South University Ave, Ann Arbor, MI 48109-1107, USA}

\author{P.~M.~Bult}
\affiliation{Department of Astronomy, University of Maryland, College Park, MD 20742, USA}
\affiliation{Astrophysics Science Division, NASA's Goddard Space Flight Center, Greenbelt, MD 20771, USA}

\author[0000-0001-7681-5845]{T.~E.~Strohmayer}
\affiliation{Astrophysics Science Division, NASA's Goddard Space Flight Center, Greenbelt, MD 20771, USA}

\author[0000-0002-6449-106X]{S.~Guillot}
\affiliation{CNRS, IRAP, 9 avenue du Colonel Roche, BP 44346, F-31028 Toulouse Cedex 4, France}
\affiliation{Universit\'{e} de Toulouse, CNES, UPS-OMP, F-31028 Toulouse, France}

\author[0000-0002-6789-2723]{G.~K.~Jaisawal}
\affil{National Space Institute, Technical University of Denmark, Elektrovej 327-328, DK-2800 Lyngby, Denmark}

\author[0000-0002-0380-0041]{C.~Malacaria}\thanks{NASA Postdoctoral Fellow}
\affiliation{NASA Marshall Space Flight Center, NSSTC, 320 Sparkman Drive, Huntsville, AL 35805, USA}
\affiliation{Universities Space Research Association, NSSTC, 320 Sparkman Drive, Huntsville, AL 35805, USA}

\author{A.~C.~Fabian}
\affiliation{Institute of Astronomy, Madingley Road, Cambridge CB3 0HA, UK}

\author[0000-0001-7681-5845]{C.~B.~Markwardt}
\affiliation{Astrophysics Science Division, NASA's Goddard Space Flight Center, Greenbelt, MD 20771, USA}

\renewcommand{\thefootnote}{\arabic{footnote}}

\begin{abstract}
We report on the first simultaneous \nicer and \nustar observations of the neutron star (NS) low-mass X-ray binary \source, obtained in 2018 August. The source was at a luminosity of $\sim1.8~(D/5.6\ \mathrm{kpc})^{2}\times10^{37}$~\lumcgs in the $0.4-30$~keV band. We account for the continuum emission with two different continuum descriptions that have been used to model the source previously. Despite the choice in continuum model, the combined passband reveals a broad Fe K line indicative of reflection in the spectrum. 
In order to account for the reflection spectrum we utilize a modified version of the reflection model \relxill\ that is tailored for thermal emission from accreting NSs. Alternatively, we also use the reflection convolution model of {\sc rfxconv} to model the reflected emission that would arise from a Comptonized thermal component for comparison. We determine that the innermost region of the accretion disk extends close to the innermost stable circular orbit (\risco) at the 90\% confidence level regardless of reflection model. 
Moreover, the current flux calibration of \nicer is within 5\% of the \nustar/FPMA(B).

\end{abstract}


\section{Introduction} \label{sec:intro}
\source is a low-mass X-ray binary (LMXB) located $5.6_{-2.1}^{+3.7}$ kpc away \citep{bailerjones18} accreting via Roche-lobe overflow from a companion star of $\sim1$~\ms.  Type-I X-ray bursts were first observed in this system in 1985 with {\it EXOSAT} \citep{vP88}, which positively identified the compact object as a neutron star (NS). The system has a mass function of $f(M)=0.53\pm0.44$~\ms\ and binary mass ratio of $q=0.05-0.41$ measured from radial velocity curves of Bowen fluorescence lines generated by X-ray irradiation of the companion star by the NS \citep{casares06}. \source has an orbital period of  $4.564\pm0.005$~hours \citep{corbet89} and is classified as an ``atoll" source based on the island-like tracks the source traces out on color-color diagrams \citep{HK89}. 

In many LMXBs, hard X-rays illuminating the accretion disk are reprocessed and reemitted in what is known as the reflection spectrum. The reprocessed continuum emission has a series of atomic features superimposed. These intrinsically narrow emission lines are broadened by special, Doppler, and general relativistic effects \citep{fabian89}. The strongest of these features is the Fe K line between $6.40-6.97$~keV. Spectral modeling of these broad reflection features are therefore used to determine properties of the NS, such as magnetic field strength \citep{cackett09b, ludlam17c}, probing the boundary layer region \citep{king16, ludlam19}, or determining the radial extent of the compact object \citep{miller13, ludlam17a}. 

The reflection spectrum of \source has been studied previously using observations taken with \chandra, \rxte, \xmm, and \bep \citep{cackett09a, torrejon10, ng10, muck13}. Simultaneous \rxte and \chandra observations were performed twice as part of a larger survey to investigate Fe lines in NS LMXBs \citep{cackett09a}. Though the \chandra gratings observations of the source did not show a clear Fe K line \citep{cackett09a, torrejon10}, the presence of a faint emission feature was not ruled out. An upper limit on the equivalent width of the Fe K line was placed at $<39.6$~eV \citep{cackett09a}. Later observations with \xmm and \bep, however showed a broad Fe line with an equivalent width between $30-56$~eV \citep{ng10, muck13}, which is consistent with the upper limit from \chandra. \source was at a similar flux level and spectral state in these different studies, so discrepancy between detection and non-detection can likely be attributed to a difference in collecting area between missions.

The {\it Nuclear Spectroscopic Telescope Array} (\nustar: \citealt{harrison13}) has provided an unhindered view of the reflection spectrum of a number of accreting NS systems (see \citealt{ludlam19} and references therein) from $3-79$~keV. The individual pixel readout of the detectors makes \nustar an ideal mission with which to observe these bright accreting binary systems devoid of pile-up effects that impact CCD-based missions. The recent installation of the {\it Neutron Star Interior Composition Explorer} (\nicer: \citealt{gendreau12}) on the {\it International Space Station} is a natural complement to \nustar\ with its 52 operational X-ray concentrator and silicon drift detector pairs that provide a large collecting area ($\sim1800$ cm$^2$ at 1.5~keV) in the $0.2-12$~keV energy band and an energy resolution of $\sim140$~eV at 6~keV. 
We obtained simultaneous observations of \source with \nicer and \nustar during 2018 August to investigate the reflection features in the combined X-ray passband. We present the observations and data reduction in \S \ref{sec:data}, the analysis and results in \S \ref{sec:results}, and conclude in \S \ref{sec:conclusion}.

\section{Observations and Data Reduction} \label{sec:data}

\nustar observed \source on 2018 August 9 starting at 21:56:09 UT. ObsID 30363003002 contains 17.6 ks of data from Focal Plane Module (FPM) A and 17.7 ks from FPMB.
The \nustar data were reduced using the standard data reduction process with {\sc nustardas} v1.8.0 and {\sc caldb} 20190314.
Spectra and \lc s are extracted using a 100$''$ radius centered on the source. Backgrounds were generated from a radial region 100$''$ away from the source. We binned the FPMA/B spectrum by a factor of 3 using {\sc grppha} \citep{choudhury17}.

\begin{figure} 
\begin{center}
\includegraphics[width=0.48\textwidth]{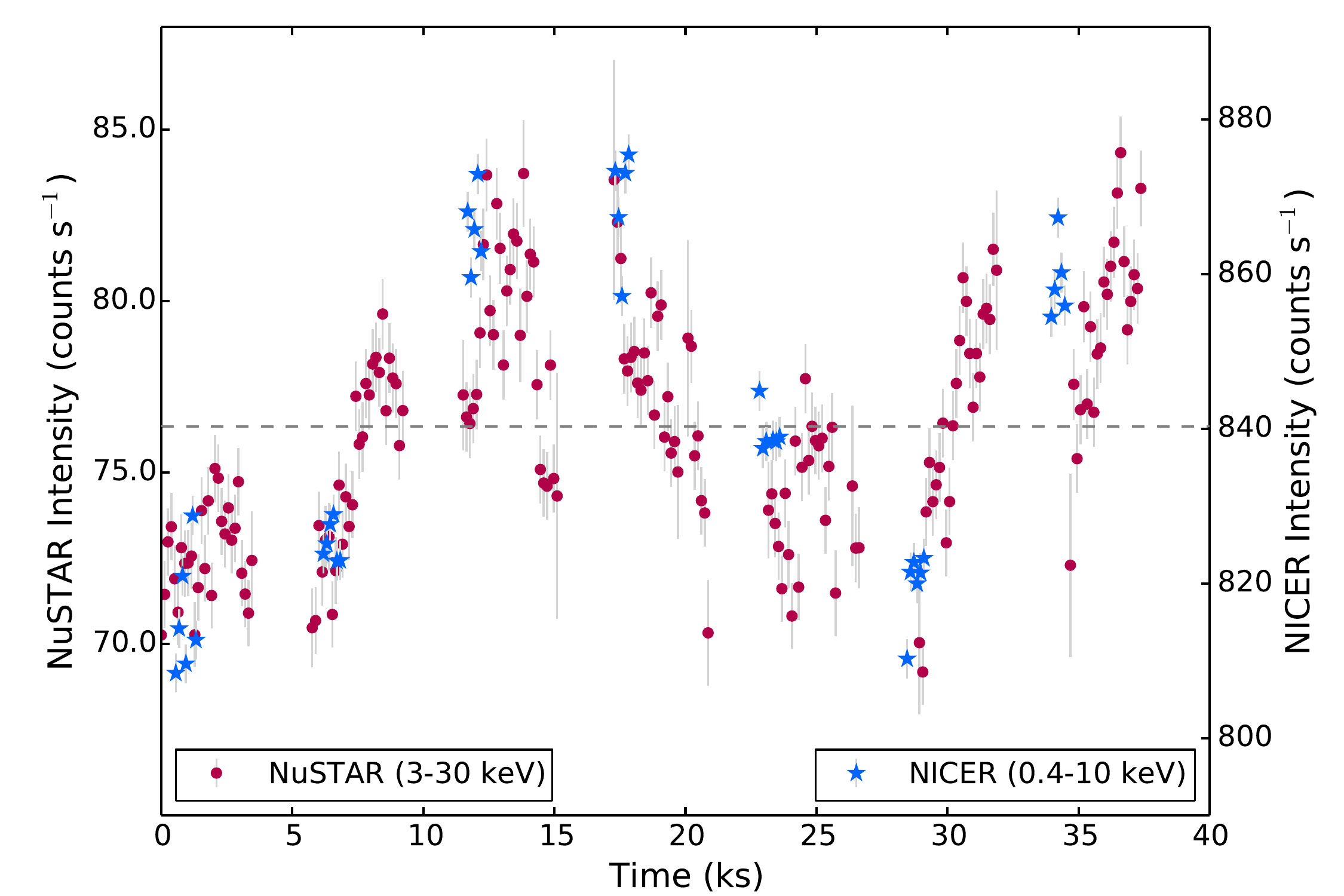}
\caption{Light-curve for the \nustar/FPMA (circles) and \nicer\ (stars) observations of \source binned to 128~s.  The grey dashed line indicates the average count rate for both \nustar\ and \nicer. 
The time elapsed is from the start of the \nustar observation on 2018-08-09 at 21:56:09UT. The source exhibits $\leq10\%$ variability over the course of the observation. Only one FPM is shown for clarity.}
\label{fig:lc}
\end{center}
\end{figure}

\begin{figure}[t!]
\begin{center}
\includegraphics[width=0.48\textwidth]{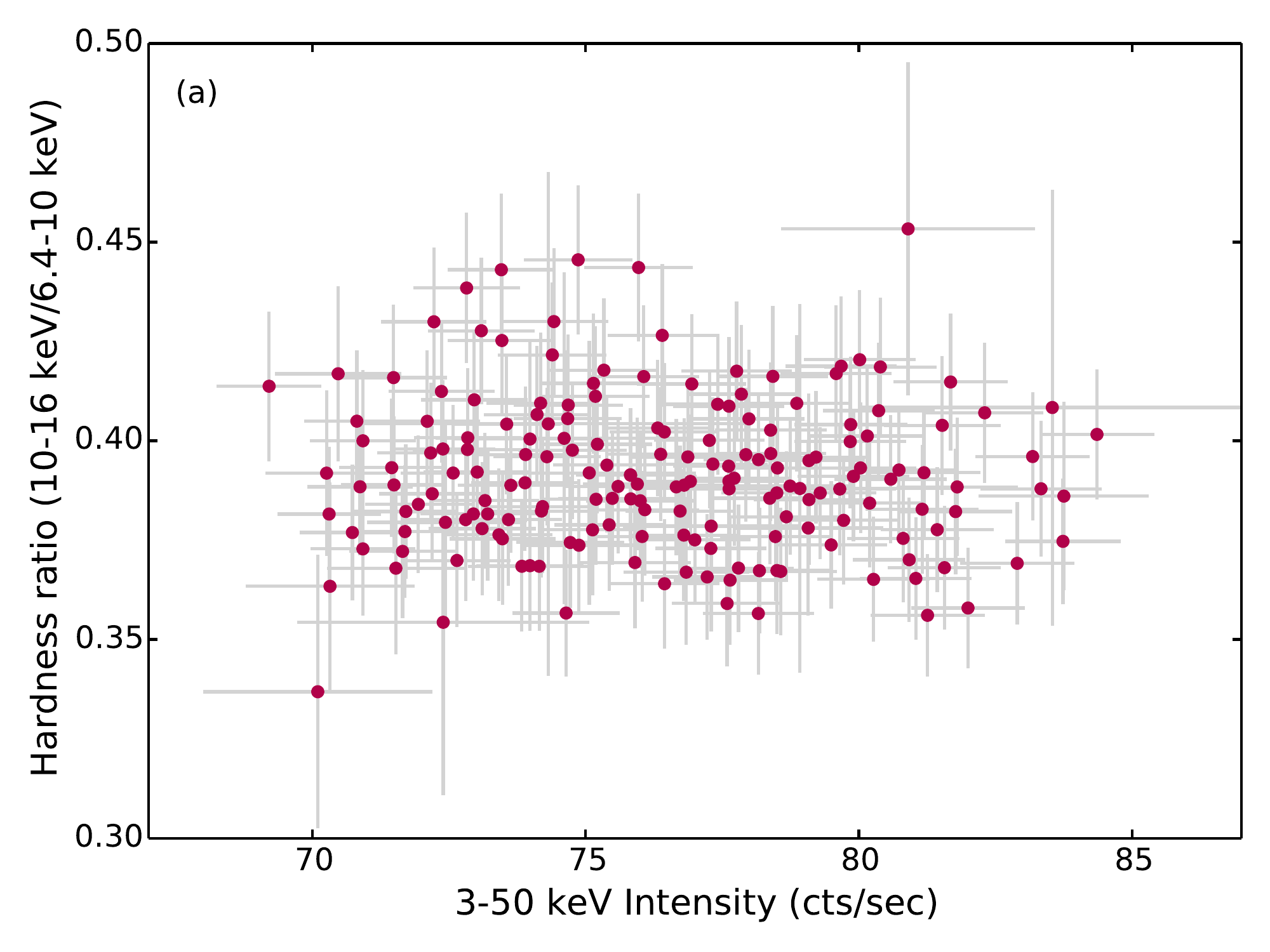}
\includegraphics[width=0.48\textwidth]{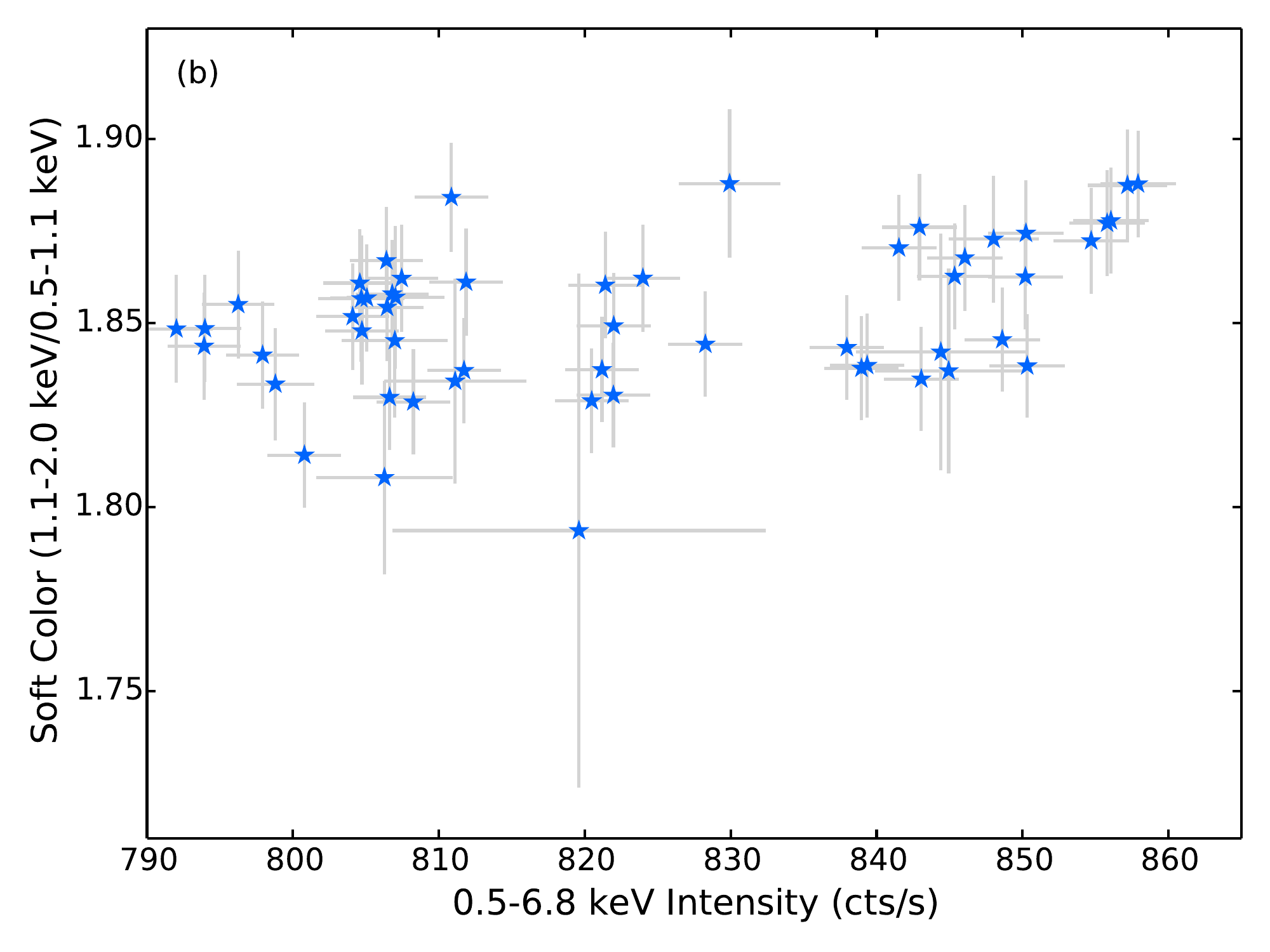}
\caption{(a) \nustar hardness versus $3-50$~keV intensity diagram and (b) \nicer soft color versus $0.5-6.8$~keV intensity diagram binned to 128~s. Although the source varies in intensity, the hardness ratio and soft color remains fairly constant. This indicates that variation in the continuum shape is minimal.}
\label{fig:HID}
\end{center}
\end{figure}

\nicer observed \source twice over the course of the \nustar observation. The first observation, ObsID 1050500103, began at 22:05:24 UT on 2018 August 9 for an exposure of 2.8 ks. The second observation, ObsID 1050500104, began at 00:00:52 UT on 2018 August 10 for 6.4 ks.  
The \nicer observations were reduced using {\sc nicerdas} 2019-06-19\_V006a. Good time intervals (GTIs) were generated using {\sc nimaketime} to select events that occurred when the particle background was low (KP~$<$~5 and COR\_SAX~$>$~4) and avoiding times of extreme optical light loading (FPM\_UNDERONLY\_COUNT~$<$~200). GTIs were applied to the data selecting PI energy channels between 25 and 1200, inclusive, and EVENT\_FLAGS=bxxx1x000 using {\sc niextract-events}. See \citet{bogdanov19} for more information on the \nicer\ screening flags. The resulting event files were loaded into {\sc xselect} to extract a combined \lc\ and spectrum. A background spectrum was generated following the same filtering criteria using {\it RXTE} ``blank sky" field~5 \citep{jahoda06}. We use the standard public RMF and the on-axis average ARF available in CALDB release 20200202 when modeling the \nicer spectrum.

No Type-I X-ray bursts were present in either data set, therefore no further processing was needed.  Figure~\ref{fig:lc} shows the \nustar/FPMA (circles) and \nicer (stars) \lc s binned to 128~s starting from when \nustar began observing \source. The source exhibits $\lesssim10\%$ variability over the $\sim40$ ks of elapsed time since the start of the observations. We check the \nustar\ hardness ratio ($10-16$~keV~/~$6.4-10$~keV: \citealt{coughenour18}) and \nicer\ soft color ($1.1-2.0$~keV~/~$0.5-1.1$~keV: \citealt{bult18}) evolution of the source (Figure~\ref{fig:HID}) and find that these remain fairly constant throughout the observation regardless of the change in intensity. This indicates that the spectral shape does not change dramatically during this time, hence we proceed with the analysis using the time averaged spectra that were extracted from the \nicer and \nustar observations.

\section{Analysis and Results} \label{sec:results}
We use {\sc xspec} v12.10.1f in our spectral analysis. \nicer data are considered in the $0.4-10$ keV energy band, while \nustar is modeled in the $3-30$ keV range as the source spectrum becomes dominated by the background at higher energies. 
There are two high bins between $3.0-3.5$~keV in the \nustar/FPMA spectrum that are instrumental in origin, but still under investigation (K.\ K.\ Madsen, $priv.\ com.$). Since it is only present in the FPMA, we chose to leave these in the analysis rather than restricting the energy range further.
To model the neutral absorption along the line of sight, we employ the {\sc xspec} model {\sc tbabs} \citep{wilms00}  with abundances set to {\sc wilm} \citep{wilms00} and {\sc vern} cross-sections \citep{verner96}. There were still two narrow absorption features present in the low energy \nicer\ spectrum. These may be astrophysical or  due to the low absorption column and luminosity of the source revealing instrumental uncertainties within the spectrum since \nicer's calibration is still ongoing. We account for these features with two {\sc edge} components. These occur at $\sim0.5$ keV and $\sim0.8$ keV, which corresponds to neutral O~K and Fe~L edges, respectively. These would be interstellar in origin if indeed they are astrophysical as opposed to instrumental.  We allow a constant to float for the FPMB and \nicer with respect to the FPMA (which is fixed at unity). 

\begin{figure} 
\begin{center}
\includegraphics[width=0.5\textwidth]{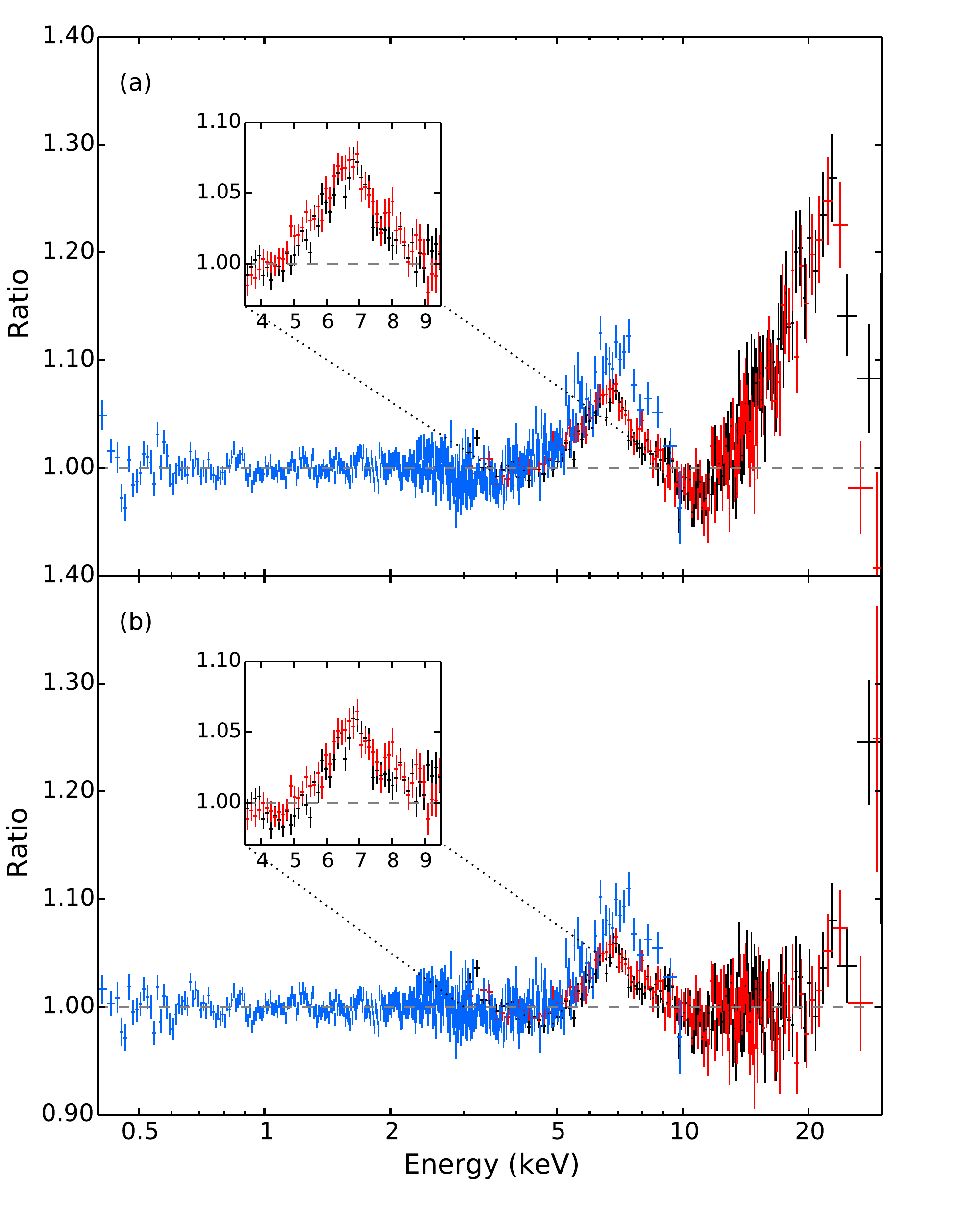}
\caption{Ratio of the \nicer (blue),  \nustar/FPMA (black), and \nustar/FPMB (red) data to the simple continuum of (a) Model 1 and (b) Model 2.  Note that the iron line region from $5–8$~keV and Compton hump region from $15-25$~keV were ignored while fitting the continuum to highlight these features and prevent them from skewing the fit shown here. The inset panel shows a close up of the \nustar Fe line profiles for each continuum description. A broad Fe K line is visible regardless of continuum choice, though the Compton hump is not as visible in panel (b) due to the high-energy rollover of the Comptonization model trying to characterize this component.}
\label{fig:ratio}
\end{center}
\end{figure}

\begin{table*}[t!]
\caption{Joint \nicer and \nustar spectral modeling of \source}
\label{tab:spectra} 
\begin{center}
\begin{tabular}{lccccccc}
\hline
Component &  Parameter & Model 1 & Model 2 & \multicolumn{2}{c}{Model 3} &  Model 4 \\
& & & & (a) Free $A_{Fe}$ &  (b) Fixed $A_{Fe}$ & \\

\hline
{\sc Constant}& $C_{\mathrm{FPMA}}$ & 1.0 & 1.0 & 1.0 & 1.0 & 1.0 \\
& $C_{\mathrm{FPMB}}$ & $0.996\pm0.002$ & $0.996\pm0.002$ & $0.995_{-0.003}^{+0.004}$ & $0.996_{-0.008}^{+0.002}$ & $0.996\pm0.002$\\
& $C_{\mathrm{NICER}}$ & $1.040\pm0.003$ & $1.039\pm0.003$ & $1.037\pm0.003$ & $1.036_{-0.013}^{+0.002}$ & $1.037_{-0.002}^{+0.004}$ \\
{\sc Tbabs} & N$_{\mathrm{H}}$ (10$^{21}$ cm$^{-2}$)& $4.91\pm0.04$ & $3.21\pm0.02$ & $4.69_{-0.04}^{+0.13}$  & $4.8_{-0.2}^{+0.3}$ &  $3.2_{-0.02}^{+0.01}$ \\
{\sc edge} & E (keV) & $0.81\pm0.01$ & $0.86\pm0.01$ & $0.82_{-0.01}^{+0.02}$ & $0.81_{-0.02}^{+0.01}$ & $0.84\pm0.01$\\
& $\tau_{\mathrm{max}}$ & $0.123\pm0.008$ & $0.044\pm0.006$ & $0.058_{-0.003}^{+0.016}$ & $0.05_{-0.01}^{+0.07}$ & $0.09\pm0.01$  \\
{\sc edge} & E (keV) & $0.526\pm0.004$ & $0.528\pm0.005$ & $0.522\pm0.004$ & $0.529_{-0.007}^{+0.002}$ &  $0.54\pm0.01$ \\
& $\tau_{\mathrm{max}}$ & $0.23\pm0.02$  & $0.11\pm0.02$ & $0.19\pm0.02$ &$0.23_{-0.01}^{+0.04}$ &  $0.208_{-0.03}^{+0.02}$\\
{\sc bbody} & $kT$ (keV) & $2.43\pm0.02$ & ...&  ... & ... & ... \\
& norm$_{bb}$ (10$^{-2}$) & $1.02\pm0.01$ & ... & ... &... & ...\\
{\sc diskbb}& $kT$ (keV)& $1.26\pm0.02$ & $0.68\pm0.01$ & $1.04\pm0.04$ & $1.03_{-0.04}^{+0.02}$ &  $0.58_{-0.02}^{+0.01}$ \\
& norm$_{disk}$ & $28_{-1}^{+2}$  & $350_{-18}^{+25}$ & $59_{-7}^{+4}$ &$58_{-11}^{+5}$&  $507_{-18}^{+38}$ \\
{\sc powerlaw} & $\Gamma$& $2.57\pm0.02$ & ... & $2.73\pm0.05$ &$2.71_{-0.11}^{+0.06}$ & ... \\
& norm$_{pl}$ & $0.45\pm0.01$ & ... & $0.30\pm0.04$ & $0.33_{-0.03}^{+0.12}$ & ...\\
{\sc nthcomp} & $\Gamma$ & ... & $1.97\pm0.03$ & ... & ... & $1.76\pm0.01$ \\
& $kT_{e}$ (keV)& ... & $3.13\pm0.04$ & ... & ... & $2.87\pm0.03$\\
& $kT_{bb}$ (keV) & ... & $1.07_{-0.04}^{+0.03}$  & ... & ... & $0.70\pm0.03$\\
& norm$_{nth}$ ($10^{-2}$) & ... &  $5.9_{-0.3}^{+0.4}$ & ... & ... &$11.3_{-0.9}^{+1.1}$\\
{\sc relxillNS} & $q$ & ... & ... & $3.3_{-1.1}^{+0.2}$ & $3.5_{-0.9}^{+0.3}$& ... \\
& $i$ ($^{\circ}$) & ... & ... & $42_{-4}^{+2}$ & $42_{-3}^{+2}$ & ...\\
& \rin\ (ISCO) & ...& ... & $1.01_{-0.01}^{+0.57}$ & $1.01_{-0.01}^{+0.74}$ & ... \\
& \rin\ (\rg) & ...& ... & $6.06_{-0.06}^{+3.42}$ & $6.06_{-0.06}^{+4.44}$ & ...\\
& $kT_{bb}$ (keV) & ... & ... & $2.89_{-0.03}^{+0.02}$ &$2.86_{-0.07}^{+0.01}$ & ... \\
& $\log(\xi)$ & ... & ... & $3.66_{-0.12}^{+0.06}$ & $3.53_{-0.12}^{+0.04}$ & ... \\
&$A_{Fe}$& ... & ... & $4.24_{-0.7}^{+1.6}$ & $2.0^{\dagger}$& ... \\
& $\log(N)$ (cm$^{-3}$) & ... & ... & $16.9_{-0.9}^{+0.4}$ & $17.0_{-0.5}^{+0.4}$ & ... \\
& $f_{refl}$ & ... & ... & $0.3\pm0.2$ & $0.3_{-0.2}^{+0.1}$ & ...\\
& norm$_{rel}$ (10$^{-3}$)& ... & ... & $1.6_{-0.5}^{+0.9}$ & $1.7_{-0.3}^{+0.1}$ & ...\\
{\sc rdblur} & Betor10 & ... &... &... &... & $-2.3_{-0.2}^{+0.1}$ \\
& \rin\ (\rg) & ... &... & ...&...& $6.10_{-0.02}^{+9.25}$ \\
& $i$ ($^{\circ}$)*& ... & ... & ... &...& $57\pm2$\\
{\sc rfxconv} & $rel_{\mathrm{refl}}$ & ... &... & ... &...& $0.19_{-0.02}^{+0.40}$\\
&$A_{Fe}$ & ... & ... & ... &...& $2.0_{-0.4}^{+0.5}$ \\
&$\cos{(i)}$* & ... & ... & ... &...& --\\
& $\log(\xi)$ & ... & ... & ... &...& $2.72_{-0.03}^{+0.05}$ \\
\hline
& $\chi^{2}/dof$ & 3106.62/1394 & 2272.66/1394 & 1661.39/1387 & 1679.12/1388 & 1735.97/1388 \\
\hline
\multicolumn{2}{l}{$^{\dagger}=\mathrm{fixed}$, $* = \mathrm{tied}$}\\
\end{tabular}
\end{center}
\medskip
Note.---  Errors are given at the 90\% confidence level. The input seed photon type in {\sc nthcomp} is set to a single temperature blackbody (inp\_type=0). 
The {\sc bbody} normalization is defined as $(L/10^{39}\ \mathrm{erg\ s^{-1}})/(D/10\ \mathrm{kpc})^{2}$. The {\sc diskbb} normalization is defined as $(R_{in}/\mathrm{km})^{2}/(D/10\ \mathrm{kpc})^{2}\times\cos{\theta}$. The power-law normalization is defined as photons keV$^{-1}$ cm$^{−2}$ s$^{−1}$ at 1~keV. The emissivity indices in {\sc relxillNS}  are tied to create a single emissivity index, $q$. The outer disk radius is fixed at a value of 990 \rg\ (165~\risco) and the dimensionless spin parameter is set to $a=0$. 
\\
\end{table*}

We fit the continuum with two models that have been previously used to describe the spectrum of \source. \citet{cackett09a} modeled the continuum with the three component model for accreting atolls as described in \citet{lin07}: a single-temperature blackbody ({\sc bbody}) for thermal emission from the NS or boundary layer, a multi-temperature blackbody ({\sc diskbb}: \citealt{mitsuda94}) to model the accretion disk, and a power-law to account for weak Comptonization. We refer to this as Model~1 in Table \ref{tab:spectra}. This model gives an  accretion disk temperature of $kT=1.26\pm0.02$~keV, single-temperature blackbody component of $kT=2.43\pm0.02$~keV, and power-law index of $\Gamma=2.57\pm0.02$. These are similar to the continuum parameters from \citet{cackett09a} when fitting the \chandra and \rxte observations.

Alternatively, \citet{muck13} described the \bep broad-band continuum with {\sc diskbb} and a thermal Comptonization component ({\sc nthcomp}: \citealt{zdziarski96, zycki99}). This is referred to as Model~2 in Table \ref{tab:spectra}.  This model returns a lower accretion disk temperature of $0.68\pm0.01$~keV, photon index of $\Gamma=1.97\pm0.03$, electron temperature of $kT_{e}=3.13\pm0.04$~keV, and seed photon temperature of $kT_{bb}=1.07_{-0.04}^{+0.03}$~keV, which agree with the range values found in \citet{muck13}.
Regardless of the choice in continuum, the presence of reflection is clearly evident in the ratio of the data to each respective model (Figure \ref{fig:ratio}), though the Compton hump at higher energies is less noticeable when using Model 2 due to the curvature of the high-energy rollover in {\sc nthcomp}.

Previous treatments of the reflected emission in this source consisted of modeling a Gaussian emission line  or {\sc diskline} component to the Fe line between $6.40-6.97$~keV \citep{cackett09a, ng10, muck13}. Here, we model the entire reflection spectrum using the special flavor of {\sc relxill} \citep{garcia14} that assumes a thermal input spectrum, $kT_{bb}$,  from the surface or boundary layer of the NS ({\sc relxillNS}) in contrast to a power-law, $\Gamma$, input of the standard model. This new model has similar parameters to \relxill\ with the addition of a variable disk density component, $\log(N \mathrm{[cm^{-3}]})$.  The other parameters of this model include an inner emissivity index ($q_{in}$), outer emissivity index ($q_{out}$), the break radius ($R_{break}$) between the two emissivity indices, dimensionless spin parameter ($a$), redshift ($z$), inclination of the system ($i$), inner disk radius (\rin) in units of the innermost stable circular orbit (ISCO), outer disk radius ($R_{out}$) in units of gravitational radii ($R_{g}=GM/c^{2}$), ionization parameter ($\log(\xi)$), iron abundance ($A_{Fe}$), and reflection fraction ($f_{refl}$).
We tie the inner and outer emissivity indices to create a single emissivity profile, $q$, making $R_{break}$ obsolete. $z=0$ since \source\ is a Galactic source. The outer disk radius is set at 990 \rg\ and the spin parameter is fixed at $a=0$ \citep{ludlam18}. We allow the reflection fraction, $f_{refl}$, to be positive so that it encompasses the single temperature blackbody from the continuum and reflection emission component. 
The results of using \relxillns\ can be found under Model 3 in Table~\ref{tab:spectra}.  

Conversely, to model the reflected emission when using the continuum description of Model 2, we use the reflection convolution model {\sc rfxconv} \citep{done06,kolehmainen11}. This generates an angle-dependent reflection spectrum from the {\sc nthcomp} input spectrum by combining the reflection emission from an ionized disk interpolated from {\sc reflionx} \citep{ross05} below 14~keV (using the average $2-10$~keV power-law index) with Compton reflected emission from {\sc pexriv} \citep{mz95} above 14~keV (using the average $12-14$~keV power-law index). The parameters of this model are the relative reflection normalization ($rel_{\mathrm{refl}}$), the Fe abundance ($A_{Fe}$), the inclination angle ($\cos{(i)}$), redshift ($z$), and ionization parameter ($\log(\xi)$).  Additionally, we convolve  {\sc rfxconv} with the relativistic blurring kernal {\sc rdblur} \citep{fabian89} to take into account general and special relativistic effects around a non-spinning compact object (i.e., $a=0$). The parameters of {\sc rdblur} are the emissivity index (Betor10: $R^{\mathrm{Betor10}}$), inner disk radius in \rg, outer disk radius, and inclination ($i$). The outer disk radius is fixed at 990 \rg\ to be consistent with \relxillns. Moreover, we tie the inclination parameters between {\sc rdblur} and {\sc rfxconv} for consistency. This is referred to as Model 4 in Table~\ref{tab:spectra}.

\begin{figure} 
\begin{center}
\includegraphics[angle=270,width=0.48\textwidth,trim=50 20 40 40, clip]{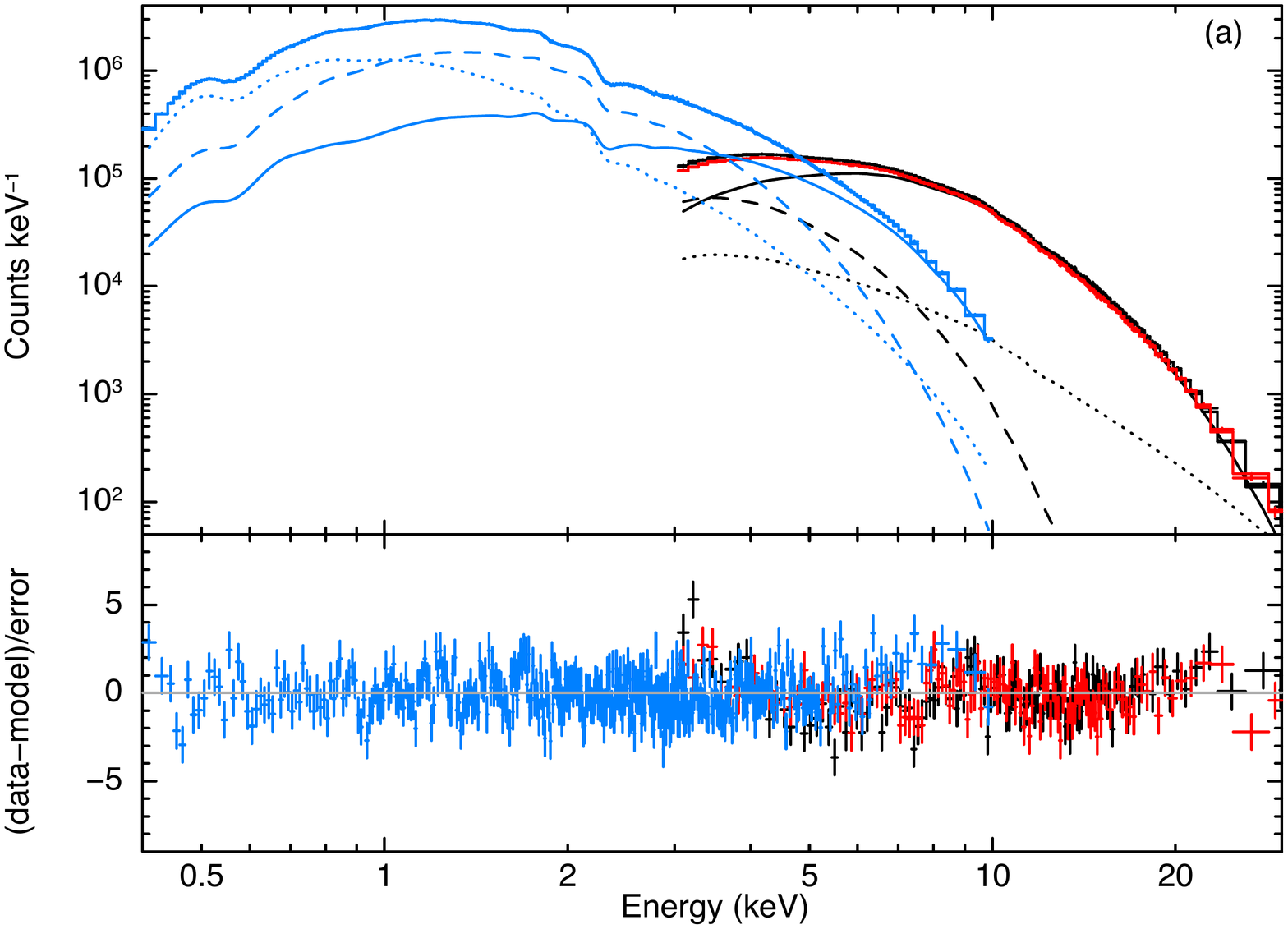}
\includegraphics[angle=270,width=0.48\textwidth,trim=50 20 40 40, clip]{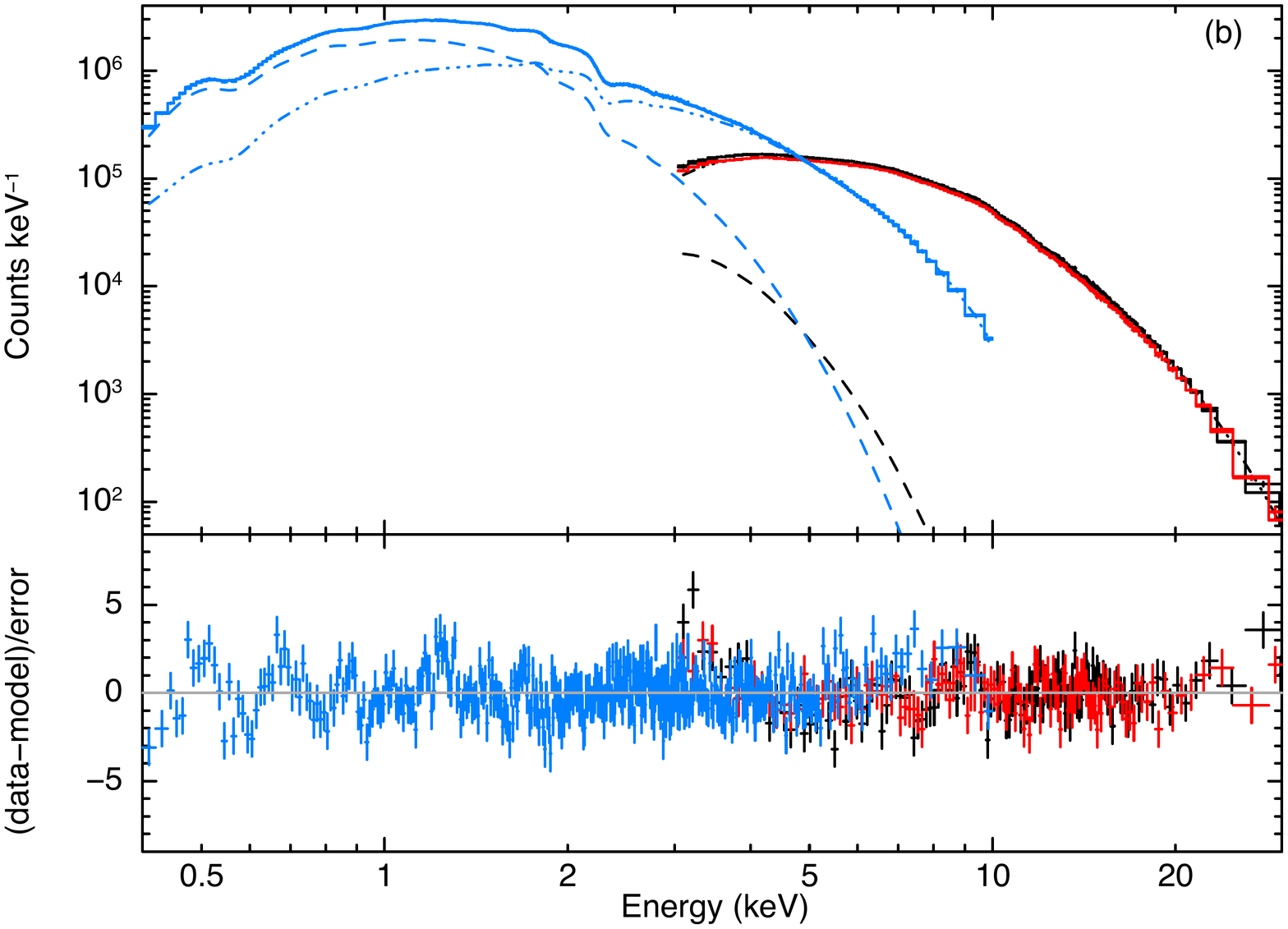}
\caption{The \nicer (blue),  \nustar/FPMA (black), and \nustar/FPMB (red) spectra and model components for (a) Model 3 (free $A_{Fe}$) and (b) Model 4 with the square-root of $\chi^{2}$ in each bin shown in the lower panels. The dashed line indicates the multi-temperature accretion disk component.  The solid line is the \relxillns\ component, which includes the single-temperature blackbody input spectrum. The dotted line shows the power-law component. The dot-dot-dot-dash line is the blurred reflection and Comptonized continuum from using {\sc rfxconv}. We only show the model components for the \nicer\ and \nustar/FPMA spectra for clarity. }
\label{fig:mocomps}
\end{center}
\end{figure}

\begin{figure} 
\begin{center}
\includegraphics[width=0.46\textwidth]{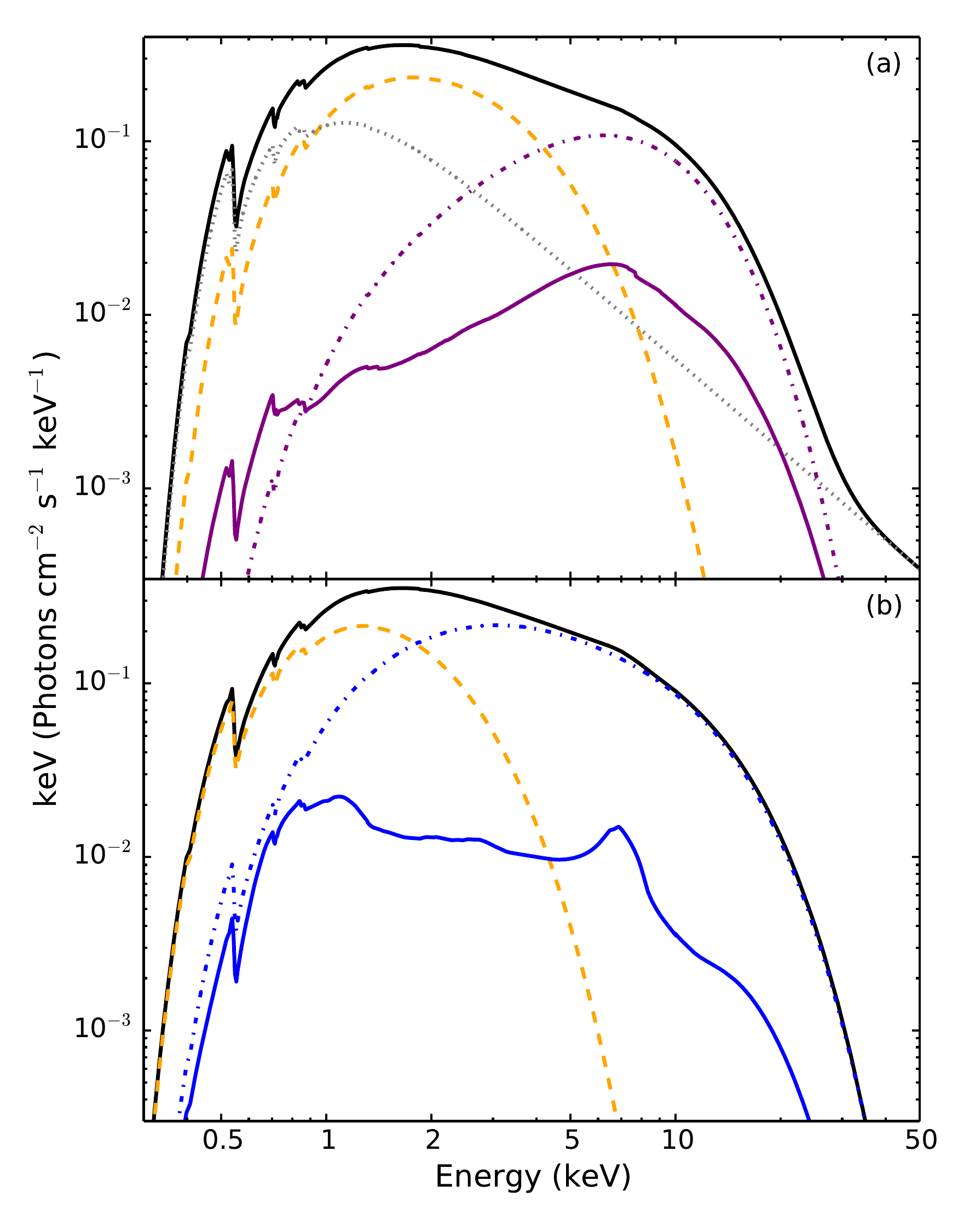}
\caption{ The unfolded model for (a) Model~3a and (b) Model~4. The solid black line indicates the overall model. The dashed orange line indicated the multi-temperature disk component in each model. The dotted grey line indicates the power-law component for Model~3a. The solid purple line in panel~(a) represents the reflection component \relxillns\ while the dot-dashed line of the same color is the input blackbody continuum component. The solid blue line in panel~(b) indicates the reflection spectrum using {\sc rfxconv} and the dot-dashed line is the input Comptonization spectrum. }
\label{fig:emocomps}
\end{center}
\end{figure}

In each case, the addition of a reflection model improves the overall fit by $\geq18\sigma$ (via an F-test) in comparison to their respective continuum model.
The multiplicative constant is within 1\% for the FPMB and within 5\% for NICER relative to the FPMA. The improvement in the \nicer\ response files are evident when compared to spectral modeling with the previous arf version nixtiaveonaxis20170601v002.arf (especially near the Au~M edges; see Figure 1 of \citealt{ludlam18}).  There is some discrepancy above 6 keV as can be seen in the differences in the blue-wing of the Fe line profile in Figure \ref{fig:ratio} and the lower panel of Figure \ref{fig:mocomps}. The reflection model fits between the difference in the shape of the Fe line profiles from \nicer\ and \nustar. The discrepancy in this region is only at the few percent level, but can contribute to uncertainty on the inclination, $i$, position of the inner disk radius, \rin, abundance of Fe, $A_{Fe}$, and disk density. It is unclear how much of this is due to \nicer's calibration, the difference in energy resolution between missions, or to the current understanding and methodology of modeling the \nicer\ background. However, it is necessary to fit the \nicer\ and \nustar\ data simultaneously. When modeling the \nicer\ data alone, the power-law index in Model 1 is not well constrained without the higher energy photons that \nustar can provide. Additionally, the emissivity index is unconstrained regardless of the continuum description when omitting the \nustar\ data. On the other hand, the \nustar spectrum is unable to constrain the neutral absorption column along the line of sight without the addition of \nicer's low-energy passband.  

The inferred neutral absorption column along the line of sight is slightly higher than the fixed value of $3.0\times10^{21}$ cm$^{-2}$ that was used in \citet{cackett09a} and \citet{muck13}.
Model~3 has an emissivity index of $q=3.3_{-0.1}^{+0.2}$, inclination of $i=42_{-4}^{+2}$~$^{\circ}$, inner disk radius of $R_{\mathrm{in}}=1.01_{-0.01}^{+0.57}$ \risco, ionization of $\log(\xi)=3.66_{-0.12}^{+0.06}$, Fe abundance (relative to solar) of $A_{Fe}=4.24_{-0.7}^{+1.6}$, disk density of $\log(N)=16.9_{-0.9}^{+0.4}$, and reflection fraction of $f_{refl}=0.3\pm0.2$.
The $A_{Fe}$ is several times higher than solar, but this is well within the range for Fe abundance that was previously reported in \citet{muck13}. To determine how this parameter impacts the values inferred from \relxillns, we fix it to a lower value of $A_{Fe}=2$. The resulting fit is $3.8\sigma$ worse than when $A_{Fe}$ was allowed to be a free parameter. However, the other parameters are all still consistent within the 90\% confidence level. Therefore, the other parameters are not highly reliant upon $A_{Fe}$. We report this fit within Table~\ref{tab:spectra} under Model 3 (b) fixed $A_{Fe}$ for direct comparison to Model 3 (a) free $A_{Fe}$.

Model 4 has an emissivity index of $-2.3_{-0.2}^{+0.1}$, inclination of $i=57\pm2^{\circ}$, inner disk radius of $R_{\mathrm{in}}=6.10_{-0.02}^{+9.25}$ \rg, ionization of $\log(\xi)=2.72_{-0.03}^{+0.05}$, Fe abundance of $A_{Fe}=2.0_{-0.4}^{+0.5}$, and reflection fraction of $rel_{refl}=0.19_{-0.02}^{+0.40}$. 
The inner disk radius agrees within the 90\% confidence level between Model 3 and 4, which indicates that the choice of continuum and reflection model does not significantly impact this result (also see \citealt{coughenour18} and \citealt{ludlam17a}). 
The discrepancy between the other recovered parameters from these two models can be attributed to the many differences between the two approaches. This is particularly true for the inclination. While the model {\sc rdblur}*{\sc rfxconv} applies an outdated relativistic convolution routine to a reflection spectrum produced from piecing together two different reflection codes ({\sc reflionx} and {\sc pexriv}), \relxillns\ self-consistently connects the angle-dependent reflection spectrum (produced with a blackbody illumination in {\sc xillverNS}) with the ray tracing code {\sc relline} \citep{dauser10}. At each disk radius, the model chooses the appropriate reflection spectrum for each emission angle calculated in a curved space-time. The resulting reflection model thus accurately captures the detailed dependence of the emitted spectrum with the viewing angle \citep{garcia14, dauser14}. Additionally, the disk density differs between these two models as Model 3 has a $\log(N)\sim17$ and Model 4 has a hard-coded density of $\log(N)=15$. A more detailed description of \relxillns\ and comparison to other reflection models is forthcoming (Garc\'{i}a, Dauser, \& Ludlam 2020, in preparation).

Figure \ref{fig:emocomps} shows the unfolded model and components for Model~3a and Model~4. The shape of the resulting reflection spectrum is very different given that the model in panel~(a) results from a blackbody input spectrum and panel~(b) is pieced together from power-law and Comptonization inputs.  
There is a flavor of {\sc relxill} where the input spectrum is a Comptonization component ({\sc relxillCP}), but this assumes that the seed photons arise from a multi-temperature blackbody (inp\_type=1 in {\sc nthcomp} for the accretion disk) which is more appropriate for black hole systems. This differs from the single-temperature input type used here and would be an inconsistent treatment of the reflection component with respect to the illuminating continuum.

\section{Conclusion} \label{sec:conclusion}
We report on the first simultaneous \nicer and \nustar observations of the persistently accreting NS LMXB \source. Regardless of the choice in continuum modeling, there are clear signatures of reflection in the \nicer\ and \nustar\ spectrum. Previous treatments of the reflection component in this system only modelled emission in the Fe line region.
We model the entire reflection spectrum using the modified version of \relxill\ that is tailored for thermal emission from a neutron star, \relxillns,  and with the reflection convolution model {\sc rfxconv}.  The source was at an unabsorbed luminosity of $1.8\pm1.1(D/5.6\ \mathrm{kpc})^{2}\times10^{37}$~\lumcgs in the $0.4-30$~keV band, which is a similar luminosity as when it was observed with \xmm, \rxte, \chandra, and \bep\ \citep{cackett09a, ng10, muck13}. This corresponds to $L/L_{\mathrm{Edd}}=0.05\pm0.03(D/5.6\ \mathrm{kpc})^{2}$ assuming a maximum Eddington luminosity of $3.8\times10^{38}$~\lumcgs \citep{kuulkers03}.

From the self-consistent \relxillns\ reflection modeling, we infer an inner disk radius of $R_{\mathrm{in}}=1.01_{-0.01}^{+0.57}$ \risco\ (Model~3a). Under the assumption of $a=0$, this corresponds to $R_{\mathrm{in}}=6.06_{-0.06}^{+3.42}$~\rg\ or $R_{\mathrm{in}}=12.52_{-0.12}^{+7.07}$~km when assuming a canonical NS mass of 1.4 \ms.  Using the alternative continuum model with {\sc rfxconv} (Model~4) provides a larger uncertainty in the inner disk radius, $R_{\mathrm{in}}=6.10_{-0.02}^{+9.25}$~\rg~$=12.61_{-0.04}^{+19.1}$~km, but is consistent within the 90\% confidence level. 
The inner disk radius inferred from the normalization of {\sc diskbb} gives a smaller but consistent range of $R_{\mathrm{in}}\sim14-15$~km in the case of Model 3a, when using a correction factor of 1.7 \citep{shimura95}, inclination of $44^{\circ}$, and $D=5.6$~kpc. The inner disk radius from {\sc diskbb} is higher for Model 4, $R_{\mathrm{in}}\sim48.5-51.2$~km, when using the same correction factor and distance, but higher inclination of $57^{\circ}$.
The inclination of the system inferred from reflection modeling differs between model,  $i=38^{\circ}-44^{\circ}$ for \relxillns\ and $i=55^{\circ}-59^{\circ}$ for {\sc rfxconv}, but both agree with the estimation from optical spectroscopy ($i=27^{\circ}-60^{\circ}$:  \citealt{casares06}). Unfortunately, the large uncertainty in the mass ratio of the system prevents our smaller range in inclination from placing meaningful constraints on the mass of the NS. 

The upper limit on the unabsorbed luminosity of \source\ (assuming $D=5.6$~ kpc) can be used to place a limit on the radial extent of a boundary layer region between the inner edge of the accretion disk and surface of the NS by using equation (25) in \citet{PS01}. The maximum extent of this region is $R_{BL}=6.1$~\rg. It is plausible that there is a boundary layer present in this system given that the inner disk radius is consistent with \risco. Conversely, we can also place an upper limit on the dipolar equatorial magnetic field strength of the NS from the upper limit on \rin.  Adapting equation (1) in \citet{cackett09b} for the magnetic dipole moment to directly provide the magnetic field strength, we obtain:
\begin{equation}
\small
\begin{aligned}
B = 3.5 \times 10^{5} \ k_{A}^{-7/4} \ x^{7/4} \left(\frac{M}{1.4\ M_{\odot}}\right)^{2} \left(\frac{10\ \mathrm{km}}{R_{NS}}\right)^3 \\ \times \left(\frac{f_{ang}}{\eta} \frac{F_{bol}}{10^{-9}\ \mathrm{erg \ cm^{-2} \ s^{-1}}}\right)^{1/2} \frac{D}{3.5\ \mathrm{kpc}}  \ \mathrm{G}
\end{aligned}
\end{equation}
Assuming an accretion efficiency of $\eta=0.2$ \citep{SS00}, unity for the conversion factor ($k_{A}$) and angular anisotropy ($f_{ang}$) as in \citet{ludlam19}, a distance of 5.6~kpc, and conical values for NS mass and radius, we acquire an upper limit of  $B\leq2.0\times10^{8}$~G from Model~3a and $B\leq4.6\times10^{8}$~G from Model~4. The magnetic field strength agrees with other accreting NS LMXBs \citep{mukherjee15, ludlam17c}. 

The current flux calibration of \nicer is within $5\%$ of the \nustar FPMA/B. Further improvements to the response files for \nicer may bring the Fe band into complete agreement with \nustar, though the current discrepancy is only at the few percent level. However, it is clear that the combined passband of \nicer and \nustar can reveal the entire reflection spectrum of these bright sources without the need to correct for pile-up effects. 
\\ 

{\it Acknowledgements:} R.M.L. thanks K.\ K.\ Madsen for discussion regarding instrumentation. Support for this work was provided by NASA through the NASA Hubble Fellowship grant \#HST-HF2-51440.001 awarded by the Space Telescope Science Institute, which is operated by the Association of Universities for Research in Astronomy, Incorporated, under NASA contract NAS5-26555. J.A.G. acknowledges support from NASA grant 80NSSC19K1020  and from the Alexander von Humboldt Foundation. E.M.C. gratefully acknowledges support through NSF CAREER award number AST-1351222.  C.M. is supported by an appointment to the NASA Postdoctoral Program at the MSFC, administered by USRA.

\facilities{ADS, HEASARC, \nicer, \nustar}
\software{heasoft (v6.26.1), nustardas (v1.8.0), nicerdas (2019-06-19\_V006a), xspec (v12.10.1)}

\end{document}